\documentclass[journal=jctcce]{achemso}
\setkeys{acs}{maxauthors=10,etalmode=truncate,chaptertitle =true,articletitle=true}
\usepackage[hidelinks]{hyperref}
\usepackage{amssymb}
\usepackage{amsmath}
\usepackage[dvipsnames]{xcolor}
\usepackage{graphicx}
\usepackage{float}
\usepackage[normalem]{ulem}
\usepackage[labelfont=bf]{caption}
\usepackage{subcaption}
\usepackage{natbib}
\usepackage{multirow}
\usepackage{bm}
\usepackage[version=4]{mhchem} 
\usepackage{attachfile2}
\usepackage{tabularx,booktabs}

\newcommand{\abs}[1]{\left|#1\right|}
\newcolumntype{L}[1]{>{\RaggedRight\arraybackslash}p{#1}}
\usepackage{ragged2e}
\usepackage{enumitem}

\newcommand{\ket}[1]{\left|#1\right\rangle}
\newcommand{\braket}[2]{\left\langle#1 |  #2\right\rangle}
\newcommand{\eri}[2]{\left\langle#1 ||  #2\right\rangle}

\newcommand{\qexpt}[3]{\left\langle#1 \left| #2 \right| #3\right\rangle}

\title{Near Chemical Accuracy From Size-Consistent Pad{\'e} Resummed M{\o}ller-Plesset Perturbation Theory}

\author{Kyle Bystrom}
\affiliation{Initiative for Computational Catalysis, Flatiron
Institute, New York, New York 10010, United States}
\author{Timothy C. Berkelbach}
\affiliation{Initiative for Computational Catalysis, Flatiron
Institute, New York, New York 10010, United States}
\altaffiliation{Columbia University,
New York, New York 10027, United States}
\author{Diptarka Hait}
\email{diptarka.hait@columbia.edu; dhait@flatironinstitute.org}
\affiliation{Initiative for Computational Catalysis, Flatiron
Institute, New York, New York 10010, United States}
\altaffiliation{Columbia University,
New York, New York 10027, United States}

\begin{document}
\maketitle
\begin{abstract} 
\noindent Pad{\'e} resummation of the correlation energy from M{\o}ller-Plesset (MP) perturbation theory is not size-consistent and is therefore seldom used. We present an orbital-based Pad{\'e} resummation of the MP2 and MP3 self-energy operators, whose trace yields a size-consistent correlation energy. We find that damping the third-order same-spin contribution leads to substantially increased accuracy, with the resulting same-spin damped orbital Pad{\'e} resummed (SSD-OPMP3) approach attaining near-chemical accuracy (mean absolute error $\sim$ 1 kcal/mol) for total atomization energies of non-multireference molecules in the W4-17 dataset against the gold standard CCSD(T) method.
Excellent performance surpassing CCSD is also obtained for other molecular datasets such as those containing barrier heights, noncovalent interactions, dipole moments, static polarizabilities, and transition-metal chemistry. SSD-OPMP3 is thus a highly accurate electronic structure method with the same noniterative $O(N^6)$ cost as MP3, filling a gap in the cost/accuracy hierarchy of wavefunction methods. 
\end{abstract}

M{\o}ller-Plesset (MP) perturbation theory\cite{moller1934note, cremer2011moller, szabo2012modern, shavitt2009many, helgaker2013molecular} is perhaps the most straightforward way to add electron correlation to a single-determinant wavefunction. Second-order MP (MP2) theory is popular both as a pure wavefunction method\cite{cremer2011moller} and as a source of orbital-based correlation for double-hybrid density functionals\cite{grimme2006semiempirical,goerigk2014double,mardirossian2018survival,shen2026high}. In contrast, higher-order MP theory is perceived to offer a suboptimal balance between cost and accuracy compared to coupled cluster (CC) methods \cite{shavitt2009many,helgaker2013molecular}.
This is largely a consequence of the poor convergence properties of the MP series, which may oscillate or diverge even for single-reference systems\cite{handy1985convergence, olsen1996surprising,christiansen1996inherent,cremer1996sixth,leininger2000mo} due to the presence of intruder states\cite{olsen2000divergence}. Third-order MP3 and fourth-order MP4 are nonetheless widely available in software packages and remain in use. 

\begin{figure} [htb!]
\begin{minipage}{0.48\linewidth}
    \centering
    \includegraphics[width=\linewidth]{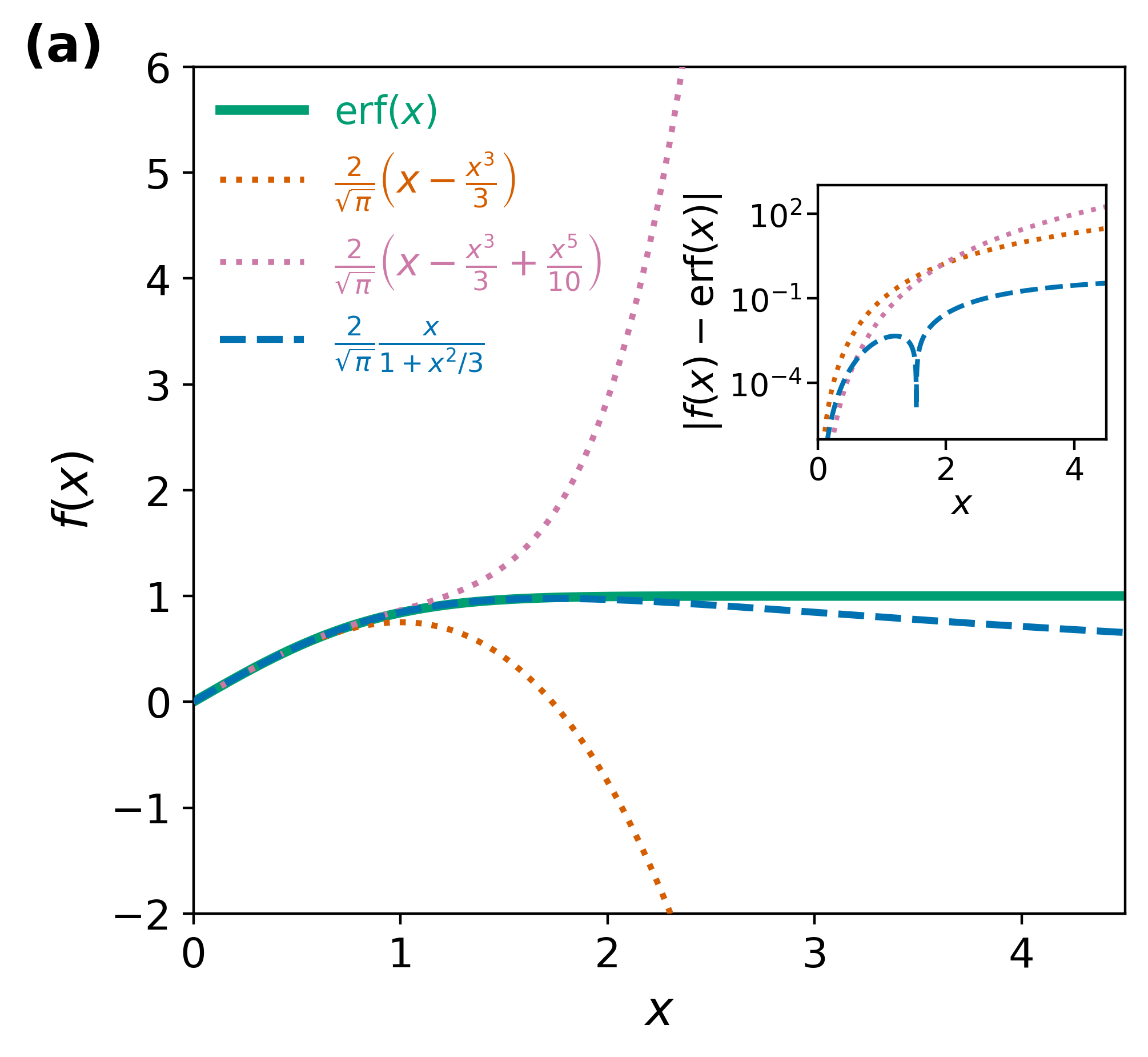}
\end{minipage}
\begin{minipage}{0.48\linewidth}
    \centering
    \includegraphics[width=\linewidth]{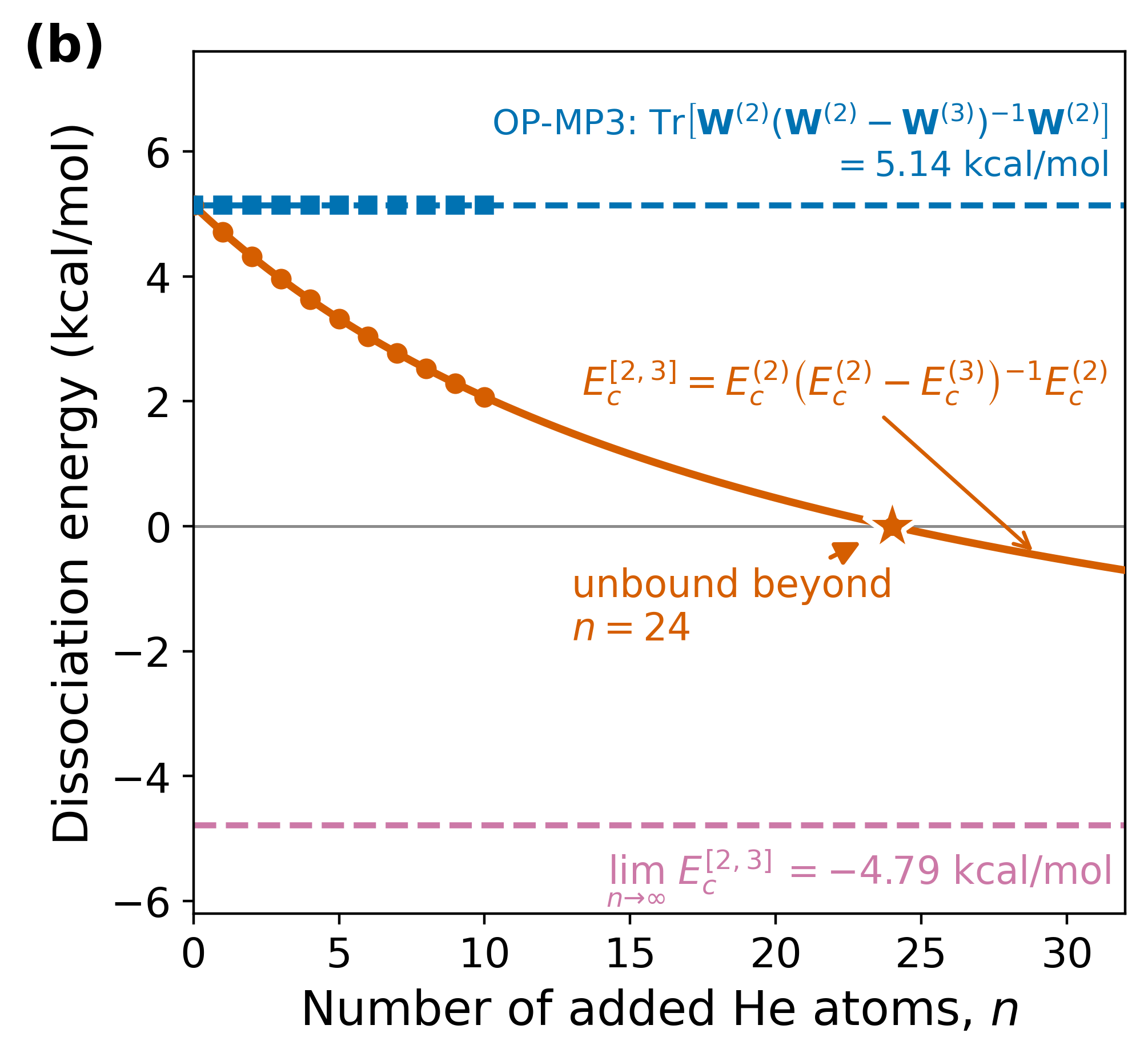}
\end{minipage}
\caption{\textbf{(a) } Comparison of Pad{\'e} resummation to truncated Taylor expansions for $\text{erf}(x)$, with the inset showing the errors for the approximations. \textbf{(b)} Demonstration of size-consistency for our approach [Eqn.~\eqref{eq:total_resum}] for the dissociation energy of a water dimer with $n$ infinitely separated He atoms, compared to scalar Pad{\'e} resummation [Eqn.~\eqref{eq:scalar_pade}]. The markers correspond to explicit calculations on supersystems with $n$ He atoms, while the curves are derived from calculations on an isolated water dimer and a single He atom. The MP calculations used HF orbitals with the aug-cc-pCVTZ basis.}
\label{fig:first}
\end{figure}

Traditional MP theory approximates electron correlation through a truncated Taylor series in the post mean-field electron-electron interaction (also known as the fluctuation potential). Truncated Taylor series are however known to be poor approximations to non-polynomial functions and are often replaced with resummations\cite{baker1961application, fischer1997use, goodson2012resummation}. Perhaps the most popular approach is Pad{\'e} resummation\cite{baker1996pade}, which uses a ratio of two polynomial functions (i.e., a rational function) with a Taylor series that matches the series of the parent function through a given order. As an illustrative example, Fig. \ref{fig:first}\textbf{(a)} compares $\text{erf}(x)=\dfrac{2}{\sqrt{\pi}}\left(x-\dfrac{x^3}{3}\ldots\right)$ to low order truncated Taylor series and the Pad{\'e} approximant $\dfrac{2}{\sqrt{\pi}}\dfrac{x}{1+\dfrac{x^2}{3}}$.

The use of resummation in MP theory has been previously explored\cite{bartlett1977comparison,laidig1985fifth,margraf2017automatic,goodson2012resummation,goodson2000convergent,zhao2024meijer,marie2021perturbation}. 
The simplest Pad{\'e} resummation of the correlation energy \cite{bartlett1977comparison,laidig1985fifth,margraf2017automatic} utilizes the traditional second- and third-order MP correlation energies $E_c^{(2)}$ and $E_c^{(3)}$ to define the resummed $E_c^{[2,3]}$ :
\begin{align}
    E_c^{[2,3]} &=   E_c^{(2)} \left(E_c^{(2)}-E_c^{(3)}\right)^{-1}E_c^{(2)} \label{eq:scalar_pade}\\
    &= \dfrac{E_c^{(2)}}{1-\dfrac{E_c^{(3)}}{E_c^{(2)}}} = E_c^{(2)} + E_c^{(3)} + \dfrac{\left(E_c^{(3)}\right)^2}{E_c^{(2)}}\ldots
\end{align}

Eqn.~\eqref{eq:scalar_pade} effectively approximates higher-order contributions via a geometric series \cite{amos1970feenberg}. Furthermore, $E_c^{[2,3]}$ is the second-order correlation energy for a Feenberg transformed zeroth order Hamiltonian\cite{feenberg1956invariance,goldhammer1956refinement,amos1970feenberg}, which yields better convergence behavior than traditional MP\cite{schmidt1993feenberg,he1996sixth,forsberg2000convergence}. 

Eqn.~\eqref{eq:scalar_pade} is nonetheless rarely used as it is not size-consistent,\cite{szabados2005size,goodson2012resummation} unlike traditional MP theory. This can be clearly illustrated for two infinitely separated arbitrary species $A$ and $B$ where the supersystem correlation energy is:
\begin{align}
    E_c^{[2,3]}[A+B] &= \dfrac{E_c^{(2)}[A\rule[3pt]{1cm}{0.4pt}B]}{1-\dfrac{E_c^{(3)}[A\rule[3pt]{1cm}{0.4pt}B]}{E_c^{(2)}[A\rule[3pt]{1cm}{0.4pt}B]}} = \dfrac{E_c^{(2)}[A] + E_c^{(2)}[B]}{1-\dfrac{E_c^{(3)}[A] + E_c^{(3)}[B]}{E_c^{(2)}[A] + E_c^{(2)}[B]}} \ne  E_c^{[2,3]}[A] + E_c^{[2,3]}[B] \label{eq:size_consistency}
\end{align}
A numerical example of the lack of size-consistency of Eqn.~\eqref{eq:scalar_pade} is shown in Fig.~\ref{fig:first}\textbf{(b)}, wherein the dissociation energy for a water dimer decreases and ultimately becomes unphysically \textit{negative} as infinitely separated He atoms are added to the supersystem. The naive Pad{\'e} correlation energy resummation in Eqn.~\eqref{eq:scalar_pade} is therefore unreliable for general chemical problems, which can often involve systems with different ratios between $E_c^{(2)}$ and $E_c^{(3)}$. However, the two sides of Eqn. \eqref{eq:size_consistency} are equal for the special case of $\dfrac{E_c^{(3)}[A]}{E_c^{(2)}[A]}=\dfrac{E_c^{(3)}[B]}{E_c^{(2)}[B]}$. This condition is trivially satisfied when $A$ and $B$ are composed of noninteracting copies of the same system.  $E_c^{[2,3]}$ for a system with $n$ noninteracting copies of a monomer is thus linear in $n$,  and  Eqn.~\eqref{eq:scalar_pade} therefore  inherits the size-extensivity of traditional MP theory\cite{laidig1985fifth}.

In this work, we report a size-consistent, Pad{\'e} resummed  MP theory by partitioning correlation energy over occupied orbitals and resumming the resulting self-energy operators. We also show that it is possible to separately resum the same-spin (SS) and opposite-spin (OS) contributions to the total correlation energy in a spin-unrestricted formalism. We find that this spin-separated Pad{\'e} resummation of the second- and third-order correlation energy leads to high accuracy when accompanied by damping of the third-order SS contribution. This same-spin damped orbital Pad{\'e} MP3 (SSD-OPMP3) method approaches the 1 kcal/mol error chemical accuracy threshold for single-reference problems with the same asymptotic cost as traditional MP3. SSD-OPMP3 therefore emerges as a highly accurate quantum chemical method with noniterative $O(N^6)$ computational cost with system size $N$.  

We now present the basics of MP theory and the route towards size-consistent resummation. 
Let $\ket{\Phi}=\ket{\Psi^{(0)}}$ be the Slater determinant that is the zeroth-order wavefunction, with $N_o$ occupied canonical spin-orbitals $i,j,k,\ldots$ and $N_v$ virtual canonical spin orbitals $a,b,c,\ldots$, which are all assumed to be real-valued for simplicity. Let $\ket{\Psi^{(n-1)}}$ be the $(n-1)^{\text{th}}$ order perturbed wavefunction from MP theory. Under the standard assumption of intermediate normalization, $\ket{\Psi^{(n-1)}}$ can be expressed as a linear combination of excitations out of $\ket{\Phi}$:
\begin{align}
    \ket{\Psi^{(n-1)}} & = \displaystyle\sum_{ia} c_{ia}^{(n-1)}\ket{\Phi_i^a} +\dfrac{1}{4}\displaystyle\sum_{ijab} c_{ijab}^{(n-1)}\ket{\Phi_{ij}^{ab}}\ldots
\end{align}
where $\ket{\Phi_i^a}$ is the Slater determinant corresponding to a single excitation from $i\to a$, $\ket{\Phi_{ij}^{ab}}$ corresponds to the double excitation from $ij\to ab$ etc., while the coefficients $c_{ia}^{(n-1)},c_{ijab}^{(n-1)}\ldots$ are completely specified by $n$th order MP theory. The $n$th order correlation energy is:
\begin{align}
    E_c^{(n)} & = \qexpt{\Psi^{(n-1)}}{\hat{H}}{\Phi}= \displaystyle\sum_{ia} c_{ia}^{(n-1)}\qexpt{\Phi_i^a}{\hat{H}}{\Phi} +\dfrac{1}{4}\displaystyle\sum_{ijab} c_{ijab}^{(n-1)}\qexpt{\Phi_{ij}^{ab}}{\hat{H}}{\Phi}
\end{align}
Triple and higher-order excitations do not contribute on account of the Slater-Condon rules\cite{szabo2012modern}. Furthermore, $\qexpt{\Phi_i^a}{\hat{H}}{\Phi}=0$ from Brillouin's theorem for an optimized Hartree-Fock (HF) solution. $E_c^{(n)}$ is thus usually written (and implemented in software packages) as:
\begin{align}
    E_c^{(n)}=\dfrac{1}{4}\displaystyle\sum_{ijab} c_{ijab}^{(n-1)}\qexpt{\Phi_{ij}^{ab}}{\hat{H}}{\Phi}=\dfrac{1}{4}\displaystyle\sum_{ijab} c_{ijab}^{(n-1)}\eri{ij}{ab}
\end{align}
with $\eri{ij}{ab}$ being the antisymmetrized two-electron integral in physicist's notation\cite{szabo2012modern}.  Defining $\Delta^{ab}_{ij}=\epsilon_a+\epsilon_b - \epsilon_i-\epsilon_j$ (where $\epsilon_p$ is the energy of the canonical spin-orbital $p$), we obtain:
\begin{align}
    c^{(1)}_{ijab}&=-\dfrac{\eri{ij}{ab}}{\Delta_{ij}^{ab}} \label{eq:c1}\\
    \implies E_c^{(2)}&=-\dfrac{1}{4}\displaystyle\sum_{ijab} \dfrac{\abs{\eri{ij}{ab}}^2}{\Delta_{ij}^{ab}}\\
    c^{(2)}_{ijab}&=-\dfrac{1}{\Delta_{ij}^{ab}}\left(\dfrac{1}{2}\displaystyle\sum_{cd} \eri{ab}{cd} c^{(1)}_{ijcd}+\dfrac{1}{2}\displaystyle\sum_{kl} \eri{kl}{ij} c^{(1)}_{klab}\right.\nonumber\\
    &\qquad\qquad\left.+\displaystyle\sum_{kc} \left[\eri{kb}{cj} c^{(1)}_{ikac}-\eri{kb}{ci} c^{(1)}_{jkac}-\eri{ka}{cj} c^{(1)}_{ikbc}+\eri{ka}{ci} c^{(1)}_{jkbc}\right]\right)\\
    \implies E_c^{(3)} &=\dfrac{1}{8}\left(\displaystyle\sum_{ijabcd} \eri{ab}{cd} c^{(1)}_{ijcd}c^{(1)}_{ijab}+\displaystyle\sum_{ijklab} \eri{ij}{kl} c^{(1)}_{klab}c^{(1)}_{ijab}-8\displaystyle\sum_{ijkabc} \eri{kb}{ic} c^{(1)}_{kjac}c^{(1)}_{ijab}\right)
\end{align}
We note that the summation over six orbital indices in $E_c^{(3)}$ is the origin of the $O(N^6)$ scaling of MP3. This is comparable to that of a single iteration of CC singles and doubles (CCSD\cite{purvis1982full}). In fact, as $N_v\gg N_o$ for any useful MP/CC calculation, the cost of both MP3 and CCSD is dominated by the $O(N_o^2N_v^4)$ scaling $\eri{ab}{cd}$ contraction. CCSD however requires many iterations at this cost to solve for the cluster amplitudes and is thus much more computationally demanding than MP3.

The lack of size-consistency of Eqn. \eqref{eq:scalar_pade} arises from $E_c^{(n)}$ being a scalar summed over the whole system.  Correlation can however be partitioned over individual occupied orbitals by:
\begin{subequations}
\begin{align}
  W_{ij}^{(n)} &= \dfrac{1}{8}\displaystyle\sum_{kab} \left(c_{ikab}^{(n-1)}\eri{jk}{ab}+c_{jkab}^{(n-1)}\eri{ik}{ab}\right) \label{eq:w_def}\\
    E_c^{(n)} &= \sum_i W^{(n)}_{ii} = \mathrm{Tr}\left[\mathbf{W}^{(n)}\right] 
\end{align}
\end{subequations}
Diagonalization of $\mathbf{W}^{(n)}$ thus yields a new set of occupied orbitals (linear combinations of canonical occupied orbitals, as defined by the eigenvectors) with associated correlation energies (eigenvalues). Quantities analogous to $\mathbf{W}^{(n)}$ have previously been reported for correlated orbital theories\cite{scuseria1995connections,bartlett2009towards}, without the perturbative indices. 
We also note that $W_{ij}^{(n)}$ is related to the two-particle one-hole block of the \textit{symmetrized} static self-energy operator\cite{szabo2012modern,kotani2007quasiparticle}. $W_{ij}^{(n)}$ also resembles pair energies in coupled electron pair approximation theories\cite{szabo2012modern,  kutzelnigg1977pair, nooijen2006orbital}, although it is important to recognize that the former represents matrix elements of a one-electron operator as opposed to a pairwise interaction energy.

A key property of $\mathbf{W}^{(n)}$ is that it is block diagonal for infinitely separated fragments $A$ and $B$,
\begin{align}
    \mathbf{W}^{(n)} = \begin{pmatrix}
    \mathbf{W}^{(n)}_{AA} &\mathbf{0}\\
    \mathbf{0} & \mathbf{W}^{(n)}_{BB}
\end{pmatrix}.
\end{align}
This follows from the size consistency of $c_{ikab}^{(n-1)}$ from MP theory, which guarantees that $c_{ikab}^{(n-1)}\eri{jk}{ab}$ must be zero when $i,j$ belong to different infinitely separated fragments. 
Matrix operations that preserve this block diagonal form can consequently be exploited to preserve size-consistency. The orbital invariance of MP theory ensures $\mathbf{W}^{(n)}$ is  invariant to rotations between virtual orbitals. A unitary transform between occupied orbitals $\mathbf{C}'_{oo}=\mathbf{U}_{oo}\mathbf{C}_{oo}$ leads to $\mathbf{W'}^{(n)}=\mathbf{U}_{oo}\mathbf{W}^{(n)}\mathbf{U}_{oo}^\dagger$ rotated by the same transformation matrix, and the trace is thus unaffected.\cite{bystrom2026size} $\mathbf{W}^{(2)}$ was in fact developed to impose size-consistency on second order Brillouin-Wigner perturbation theory\cite{carter2023repartitioned,coveney2023regularized, dittmer2025repartitioning,shen2026improved} and has also been utilized to develop size-consistent adiabatic connection-based functionals\cite{bystrom2026size}. The use of $\mathbf{W}$ for size-consistent third-order Brillouin-Wigner perturbation theory has very recently been reported\cite{wang2026thirdsc}.

We define size-consistent Pad{\'e} resummed correlation models through matrix operations on $\mathbf{W}^{(2)}$ and $\mathbf{W}^{(3)}$. For notational convenience, we define a matrix function $\mathcal{P}$:
\begin{align}
    \mathcal{P}(\mathbf{A},\mathbf{B}) &= \mathbf{A}(\mathbf{A}-\mathbf{B})^{-1}\mathbf{A}  \label{eq:pdef}
\end{align}
The size-consistent orbital Pad{\'e} MP3 (OP-MP3) correlation energy is thus:
\begin{align}
    E_c^{\text{OP-MP3}} &=\mathrm{Tr}\left[ \mathcal{P}(\mathbf{W}^{(2)},\mathbf{W}^{(3)}) \right] = \mathrm{Tr}\left[\mathbf{W}^{(2)}(\mathbf{W}^{(2)}-\mathbf{W}^{(3)})^{-1}\mathbf{W}^{(2)}\right] \label{eq:total_resum}
\end{align}
Size-consistency (and thus, size-extensivity) naturally follows from $\mathbf{W}^{(2)}$ and $\mathbf{W}^{(3)}$ being block-diagonal in noninteracting fragments, which causes matrix algebra to be local to the corresponding blocks. Furthermore, Eqn. \eqref{eq:total_resum} is invariant to occupied-occupied and virtual-virtual rotations, like MP theory. Fig. \ref{fig:first}\textbf{(b)} provides numerical evidence for a simple model system where the scalar Eqn. \eqref{eq:scalar_pade} is not size-consistent. 
The computational cost remains dominated by the same $O(N_o^2N_v^4)$ scaling evaluation of $c_{ikab}^{(2)}$ as in traditional MP3. Subsequent construction of $\mathbf{W}^{(3)}$ scales as $O(N_o^3N_v^2)$ and further matrix operations as $O(N_o^3)$. There is thus no change in asymptotic scaling compared to traditional MP3. We further note that we have not found a partitioning of the Hamiltonian for which Eqn. \eqref{eq:total_resum} is the second order perturbation energy, unlike the role played by the Feenberg transform for the scalar resummation in Eqn. \eqref{eq:scalar_pade}.  

Eqn. \eqref{eq:total_resum} utilizes 100\% of the third-order correlation energy. However, the most successful MP3-based methods at present are the so-called ``MP2.$X$" approaches\cite{pitovnak2009scaled,riley2011mp2,bertels2019third,rettig2020third,loipersberger2021exploring,wang2026third} where the correlation energy $E_c^{\text{MP2.}X}=E_c^{(2)}+xE_c^{(3)}$ only uses a fraction $x=X/10$ of the third-order correlation energy. Perhaps the most well-known member of the family is MP2.5\cite{pitovnak2009scaled}, which averages the total correlation energies from the traditional MP2 and MP3 methods. The second-order contribution $E_c^{(2)}$ is thus unscaled, allowing this approach to correctly describe the asymptotic limit of dispersion interactions\cite{lochan2005scaled}. More recently, models with $x\sim 0.8$ that use density functional theory (DFT) or $\kappa$-regularized orbital-optimized MP2\cite{lee2018regularized} ($\kappa$OOMP2) Slater determinants have been reported to provide excellent performance on standard quantum chemical benchmark sets\cite{bertels2019third,rettig2020third}.  The empirical success of MP2.$X$ models thus motivates the parameterized OP-MP3$(x)$ method with correlation energy:
\begin{align}
	E_c^{\text{OP-MP3}}(x) &= \mathrm{Tr}\left[ \mathcal{P}(\mathbf{W}^{(2)},x\mathbf{W}^{(3)}) \right]
     \label{eq:total_resum_scale}
\end{align}
where the third order $\mathbf{W}^{(3)}$ is scaled by a constant factor $x$. OP-MP3 without an explicit specification of $x$ will default to the case of $x=1$ [Eqn. \eqref{eq:total_resum}]. 

Eqn.~\eqref{eq:w_def} uses double-excitation configuration interaction coefficients. For spin-unrestricted orbitals, it is possible to separate the contributions arising from same-spin (SS) and opposite-spin (OS) double excitations as follows:
\begin{align}
\mathbf{W}^{(n)} &= \begin{pmatrix}
    \mathbf{W}^{(n)}_{\alpha\alpha} &\mathbf{0}\\
    \mathbf{0} & \mathbf{W}^{(n)}_{\beta\beta}\end{pmatrix} = \begin{pmatrix}
    \mathbf{W}^{\text{SS}(n)}_{\alpha\alpha} + \mathbf{W}^{\text{OS}(n)}_{\alpha\alpha} &\mathbf{0}\\
    \mathbf{0} & \mathbf{W}^{\text{SS}(n)}_{\beta\beta} +  \mathbf{W}^{\text{OS}(n)}_{\beta\beta}\end{pmatrix}\\
    W_{i_{\sigma}j_{\sigma}}^{\text{SS}(n)} &= \dfrac{1}{8}\displaystyle\sum_{k_\sigma a_\sigma b_\sigma} \left(c_{i_\sigma k_\sigma a_\sigma b_\sigma}^{(n-1)}\eri{j_\sigma k_\sigma}{a_\sigma b_\sigma}+c_{j_\sigma k_\sigma a_\sigma b_\sigma}^{(n-1)}\eri{i_\sigma k_\sigma}{a_\sigma b_\sigma}\right) \label{eq:wss_def}\\
    W_{i_{\sigma}j_{\sigma}}^{\text{OS}(n)} &= \dfrac{1}{4}\displaystyle\sum_{k_{\bar{\sigma}} a_\sigma b_{\bar{\sigma}}} \left(c_{i_\sigma k_{\bar{\sigma}} a_\sigma b_{\bar{\sigma}}}^{(n-1)}\braket{j_\sigma k_{\bar{\sigma}}}{a_\sigma b_{\bar{\sigma}}}+c_{j_\sigma k_{\bar{\sigma}} a_\sigma b_{\bar{\sigma}}}^{(n-1)}\braket{i_\sigma k_{\bar{\sigma}}}{a_\sigma b_{\bar{\sigma}}}\right) \label{eq:wos_def}
\end{align}
where $\sigma\in\{\alpha,\beta\}$ and $\bar{\sigma}$ denotes the opposite spin to $\sigma$.  We note that $W_{i_{\sigma}j_{\bar{\sigma}}}^{(n)}=0$ follows directly from Eqn.~\eqref{eq:w_def}, preventing any unphysical coupling between spin blocks. It is now possible to define resummed same-spin (SS) and opposite-spin (OS) correlation energy models:
\begin{align}
    E_c^{\text{OP-MP3(SS)}}(x_{ss}) =& \mathrm{Tr}\left[\mathcal{P}\left(\mathbf{W}_{\alpha\alpha}^{\text{SS}(2)},x_{ss}\mathbf{W}_{\alpha\alpha}^{\text{SS}(3)}\right)\right] + \mathrm{Tr}\left[\mathcal{P}\left(\mathbf{W}_{\beta\beta}^{\text{SS}(2)},x_{ss}\mathbf{W}_{\beta\beta}^{\text{SS}(3)}\right)\right]
    \label{eq:ss_resum_scale}\\
    E_c^{\text{OP-MP3(OS)}}(x_{os}) =&      \mathrm{Tr}\left[\mathcal{P}\left(\mathbf{W}_{\alpha\alpha}^{\text{OS}(2)},x_{os}\mathbf{W}_{\alpha\alpha}^{\text{OS}(3)}\right)\right] + \mathrm{Tr}\left[\mathcal{P}\left(\mathbf{W}_{\beta\beta}^{\text{OS}(2)},x_{os}\mathbf{W}_{\beta\beta}^{\text{OS}(3)}\right)\right]
    \label{eq:os_resum_scale}
\end{align}
where $x_{ss}$ and $x_{os}$ are scaling parameters for the same-spin and opposite-spin third-order correlation energies. The SS and OS resummed correlation energies can be combined to yield the general spin-separated orbital Pad{\'e} MP3 (SOP-MP3) correlation model:
\begin{align}
    E_c^{\text{SOP-MP3}}(x_{ss},x_{os})&= E_c^{\text{OP-MP3(SS)}}(x_{ss}) + E_c^{\text{OP-MP3(OS)}}(x_{os})  \label{eq:sop_resum_scale}
\end{align}

The SS and OS correlation energies behave distinctly in chemical systems, e.g.,  showing different behavior with respect to angular momentum for atomic systems, due to the Fermi-hole present for same-spin interactions\cite{schwartz1962importance}. This leads to different convergence rates of SS and OS correlation against basis set size\cite{petersson1985complete,klopper2001highly}. 
Separate resummation of the SS and OS correlation energies via Eqns.~\eqref{eq:ss_resum_scale}--\eqref{eq:sop_resum_scale} thus appears to be more physically motivated than Eqn.~\eqref{eq:total_resum}. 

The empirical success of spin-component scaled (SCS) methods\cite{grimme2003improved,grimme2003improvedmp3,jung2004scaled,grimme2012spin} motivates the presence of $x_{ss}$ and $x_{os}$. We however note that Eqn. \eqref{eq:sop_resum_scale} fundamentally differs from the previously defined SCS-MP3 method\cite{grimme2003improvedmp3} in two important ways. SCS-MP3 applies spin-component scaling to the \textit{second-order} SS and OS correlation energies, while only applying a uniform spin-agnostic scaling to $E_c^{(3)}$. In contrast, Eqn. \eqref{eq:sop_resum_scale} does not alter the second order contribution in any way (preserving correct long range asymptotic behavior for dispersion) and aims to permit the use of different scaling for the \textit{third-order} SS and OS contributions. SOP-MP3 without explicit specification of scaling parameters will default to $x_{ss}=x_{os}=1$, for brevity. 

The utility of Eqns.  \eqref{eq:total_resum_scale}  and \eqref{eq:sop_resum_scale} for chemical problems is now assessed through their performance on benchmark datasets\cite{liang2025gold}. It has been previously shown that the performance of MP3 is significantly degraded when HF orbitals are used\cite{bertels2019third,rettig2020third,loipersberger2021exploring,wang2026third} and we therefore focus on reference Slater determinants generated from $\kappa$OOMP2 and DFT. The references were processed in the manner described in Ref. \citenum{rettig2020third} (fully elaborated in the Computational Methods section), which ensures that $\mathbf{W}^{(n)}$ utilized are free from any explicit dependence on DFT orbital energies. We compare the MP methods to the coupled cluster singles and doubles with perturbative triples [CCSD(T)] method\cite{raghavachari1989fifth} with HF orbitals, which is considered to be the gold standard for single-reference quantum chemistry. All calculations are carried out in the aug-cc-pCVTZ basis\cite{dunning1989gaussian,kendall1992electron,woon1993gaussian,woon1995gaussian,peterson2002accurate,prascher2011a, deyonker2007systematically} without any electrons held frozen, unless specified otherwise.

\begin{figure}[htb!]
    \centering
    \includegraphics[width=\linewidth]{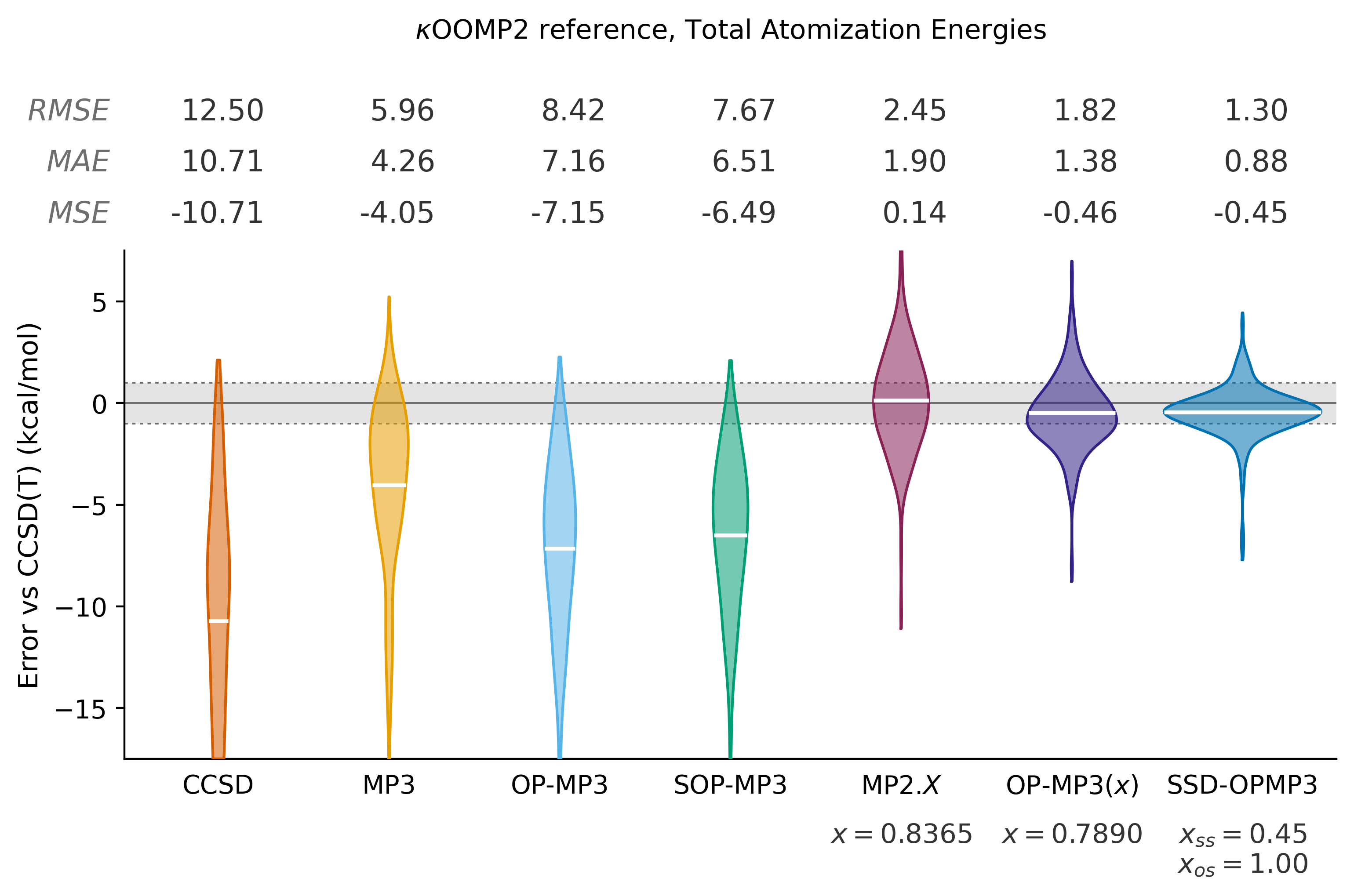}
    \caption{Distribution of errors (as kernel density estimates) in total atomization energies relative to
  CCSD(T) for the 183 molecules in the nonMR subset of W4-17. The MP
  methods use $\kappa$OOMP2 optimized Slater determinants as the zeroth order wavefunction, while CCSD and the CCSD(T) reference use
  HF orbitals.  RMSE, MAE and MSE (kcal/mol) are
  reported above the plot, with the white bars marking the
  MSE. The shaded band marks $\pm1$ kcal/mol about the CCSD(T) benchmark, and the scaling parameters are given below each label.  All calculations are in the aug-cc-pCVTZ basis with no frozen core. }
    \label{fig:w417errors}
\end{figure}

We begin by considering the non-multireference (nonMR) subset of the W4-17 set of total atomization energies (TAEs) of small main-group molecules\cite{karton2017w4}. The performance of various MP methods with $\kappa$OOMP2 orbitals, as well as CCSD with HF orbitals, is shown in Fig. \ref{fig:w417errors}. Traditional MP3 with $\kappa$OOMP2 orbitals is significantly more accurate than CCSD, having roughly half the RMSE. Systematic underbinding is the main source of error for both approaches. Unscaled OP-MP3 yields disappointing performance, underbinding to a greater extent than traditional MP3. Spin-separated resummation in SOP-MP3 only marginally improves performance compared to OP-MP3.

Least-squares fitting to TAEs reveals that the parameterized models are capable of yielding significantly better performance. The optimal MP2.$X$ model ($x=0.8365$) yields an RMSE of 2.45 kcal/mol and virtually no systematic error, substantially improving upon unscaled MP3. Fitting the SS and OS third-order contributions separately \textit{without} resummation yields little further improvement in performance (RMSE of 2.37 kcal/mol), although the optimal $x_{os}=0.9876$ is interestingly very close to $1$, while SS third-order correlation is damped by $x_{ss}=0.6499$. The combination of resummation and scaling proves to be considerably more effective, as OP-MP3 with $x=0.7890$ yields an RMSE of 1.82 kcal/mol. SOP-MP3 with $x_{ss}=0.3809$ and $x_{os}=1.0253$ is even more effective, yielding an RMSE of 1.25 kcal/mol. The optimal $x_{os}$ value motivates us to fix $x_{os}=1$ and only fit $x_{ss}$, obtaining a negligible ($<0.01$ kcal/mol) increase in RMSE for $x_{ss}=0.4279$. Rounding to $x_{ss}=0.45$ and keeping $x_{os}=1.0$ yields an RMSE of 1.30 kcal/mol (MSE $-$0.45 kcal/mol). While the performance of SOP-MP3($x_{ss}=0.45, x_{os}=1$) is thus measurably worse than the performance of canonical CCSD(T) itself (the combined post-CCSD(T) contributions for valence electrons in Ref \citenum{karton2017w4} indicate an RMSE of 0.46 kcal/mol for this dataset),
it far surpasses CCSD, traditional MP3 and the optimal MP2.$X$ model. We therefore define same-spin damped orbital Pad{\'e} MP3 (SSD-OPMP3) as a special case of Eqn. \eqref{eq:sop_resum_scale} with the correlation energy given by: 
\begin{align}
    E_c^{\text{SSD-OPMP3}}(x_{ss})&= E_c^{\text{OP-MP3(SS)}}(x_{ss}) + E_c^{\text{OP-MP3(OS)}}(1.00)  \label{eq:ssd_resum_scale}
\end{align}

\begin{figure}[htb!]
    \centering
    \includegraphics[width=\linewidth]{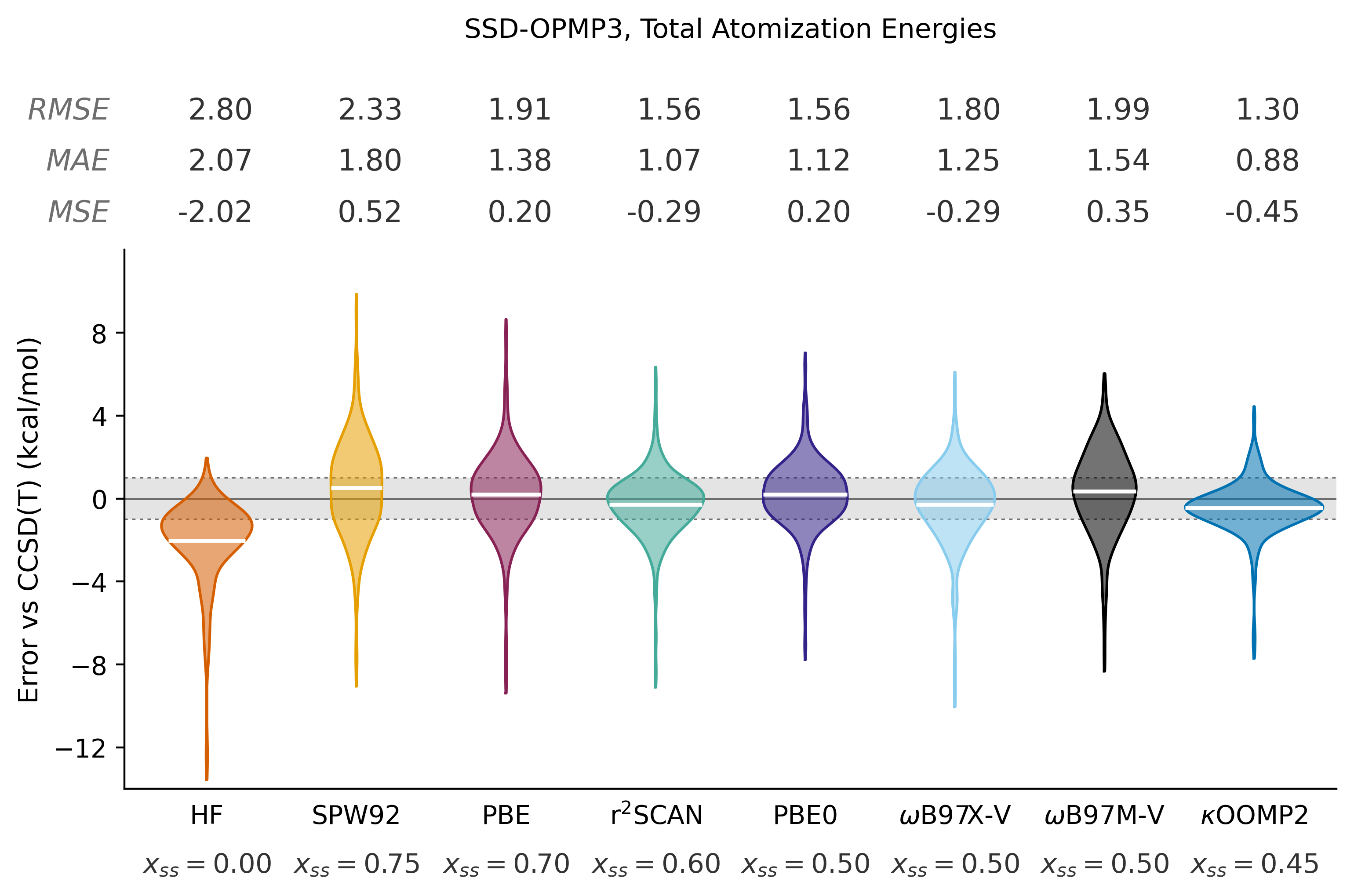}
    \caption{Distribution of errors in total atomization energies relative to
  CCSD(T) for the 183 molecules of the nonMR subset of W4-17, for SSD-OPMP3 on different references. The same-spin damping parameters $x_{ss}$ are noted below each label. All calculations are in the aug-cc-pCVTZ basis with no frozen core. }
    \label{fig:w417_orbitals}
\end{figure}

The mean absolute error (MAE) of 0.88 kcal/mol for SSD-OPMP3 with a $\kappa$OOMP2 reference and $x_{ss}=0.45$ can be viewed as an indication that this method has effectively attained the popular 1 kcal/mol threshold for chemical accuracy. A claim of ``near" chemical accuracy is perhaps more appropriate as comparisons are being made to CCSD(T) and not full configuration interaction, and a nontrivial proportion (31\%) of the TAE errors lies outside the 1 kcal/mol window. On the other hand, TAEs are a rather difficult prediction target as the errors accumulate with system size, and the performance of SSD-OPMP3 is quite impressive on the whole. Interestingly, ``correcting" the reference CCSD(T)/aug-cc-pCVTZ TAEs by the post-CCSD(T) contributions from smaller basis sets that are reported in Ref \citenum{karton2017w4} gives an SSD-OPMP3 RMSE of 1.33 kcal/mol and MAE of 0.95 kcal/mol. There does not appear to be a strong relationship between the post-CCSD(T) contributions and the extent to which SSD-OPMP3 deviates from CCSD(T), at least for this nonMR dataset.

Fig. \ref{fig:w417_orbitals} shows that the use of Slater determinants obtained from DFT yields similar, if slightly worse behavior compared to CCSD(T) for the nonMR subset of W4-17. The reported $x_{ss}$ corresponds to the optimal fit for TAE RMSE minimization with Eqn. \eqref{eq:ssd_resum_scale} rounded to the closest multiple of 0.05, with two exceptions. The optimal fit for HF has $x_{ss}=-0.1643 < 0$, and $0$ is chosen instead, while the $\omega$B97X-V\cite{mardirossian2014omegab97x} fit is $x_{ss}=0.4702$ but $0.5$ is chosen in order to associate all hybrid functionals with the same value. Rather interestingly, there is perceptible variation in performance vs the choice of the reference Slater determinant, unlike what was previously observed for MP2.$X$\cite{rettig2020third}. The best results appear to arise from $\kappa$OOMP2, PBE0\cite{pbe0}, and r$^2$SCAN\cite{furness2020accurate} optimized Slater determinants, while HF is not unexpectedly the worst performer. We focus on $\kappa$OOMP2 and PBE0 references for the most part in this work, but results from HF, SPW92\cite{Slater,PW92}, PBE\cite{PBE}, r$^2$SCAN, $\omega$B97X-V, and $\omega$B97M-V\cite{mardirossian2016omegab97m} are provided in the supporting information. We also note that subsequent references to SSD-OPMP3 in this work for a given reference utilize the $x_{ss}$ defined in Fig. \ref{fig:w417_orbitals}.

\begin{figure}[htb!]
    \includegraphics[width=\linewidth]{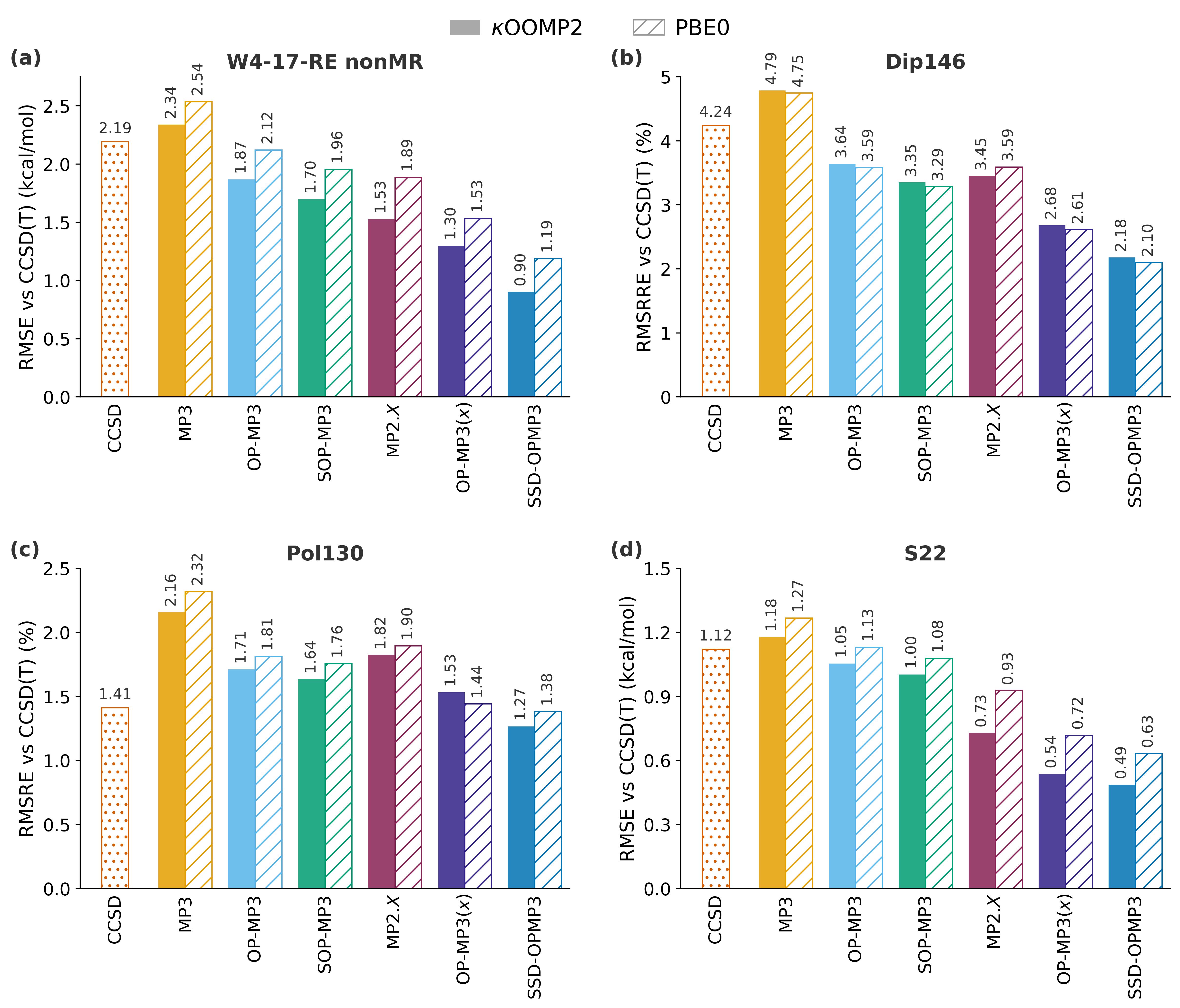}
\caption{Performance of CCSD and the various MP models against CCSD(T) in the same basis for (a) 22769 possible reactions in the W4-17-RE dataset\cite{morgante2019accdb} that only involve nonMR molecules, (b) dipole moments in the Dip146 dataset, (c) static polarizabilities in the Pol130 dataset, and (d) noncovalent interactions (all counterpoise corrected) in the S22 dataset. Solid bars denote $\kappa$OOMP2 references while striped bars correspond to PBE0 references for the MP methods (CCSD uses HF orbitals throughout). 
(a) and (d) report RMSE in kcal/mol, (b) reports RMS regularized relative error (RMSRRE) \cite{hait2018accurate} in \% and (c) reports RMS relative error (RMSRE) in \%.}
\label{fig:pair_errors}
\end{figure}

The good performance of SSD-OPMP3 over the nonMR W4-17 TAE training set does not necessarily mean that damping SS correlation is generally appropriate for chemical problems. Fig. \ref{fig:pair_errors} therefore reports the performance of the MP models discussed in this work for four other benchmark datasets: (a) 22769 possible reactions in the W4-17-RE dataset\cite{morgante2019accdb} that only involve single reference molecules; (b) the 145\footnote{Ref. \citenum{liang2025gold} only has 145 dipoles despite the name of this dataset. The ``missing" molecule is triplet \ce{CH2}.} molecular dipole moments in the Dip146 dataset\cite{hait2018accurate}; (c) the static polarizabilities in the Pol130 dataset\cite{hait2018accuratepolar} and (d) the 22 noncovalent interactions in the S22 dataset\cite{jurevcka2006benchmark}. We observe that performance improves perceptibly on moving from traditional MP3 to OP-MP3 for all four datasets, indicating the efficacy of orbital Pad{\'e} resummation. There is a rather slight decrease in error for SOP-MP3 vs OP-MP3, but same-spin damping in SSD-OPMP3 leads to a very significant reduction in error, resulting in better performance than CCSD for all four datasets. This offers strong evidence for the efficacy of SSD-OPMP3 as a highly accurate $O(N^6)$ scaling approach. Remarkably, SSD-OPMP3 attains an RMSE of 0.9 (1.2) kcal/mol vs CCSD(T) with $\kappa$OOMP2 (PBE0) references for the nonMR W4-17-RE reaction energies, which compares very well against the chemical accuracy threshold. This dataset is nonetheless related to the training set, as all reaction energies can be expressed as differences in TAEs of the W4-17 nonMR molecules.

In contrast, the dipole moments and polarizabilities are much more chemically distinct test sets and the good performance of SSD-OPMP3 relative to SOP-MP3 offers direct support for damping same-spin third-order correlation in orbital Pad{\'e} resummed models. The improved performance in predicting electrical response properties like dipole moments and static polarizabilities furthermore indicates efficacy in modeling noncovalent interactions, which is supported by the excellent behavior of SSD-OPMP3 for the S22 dataset (Fig. \ref{fig:pair_errors}(d)). 

\begin{figure}[htb!]
    \includegraphics[width=\linewidth]{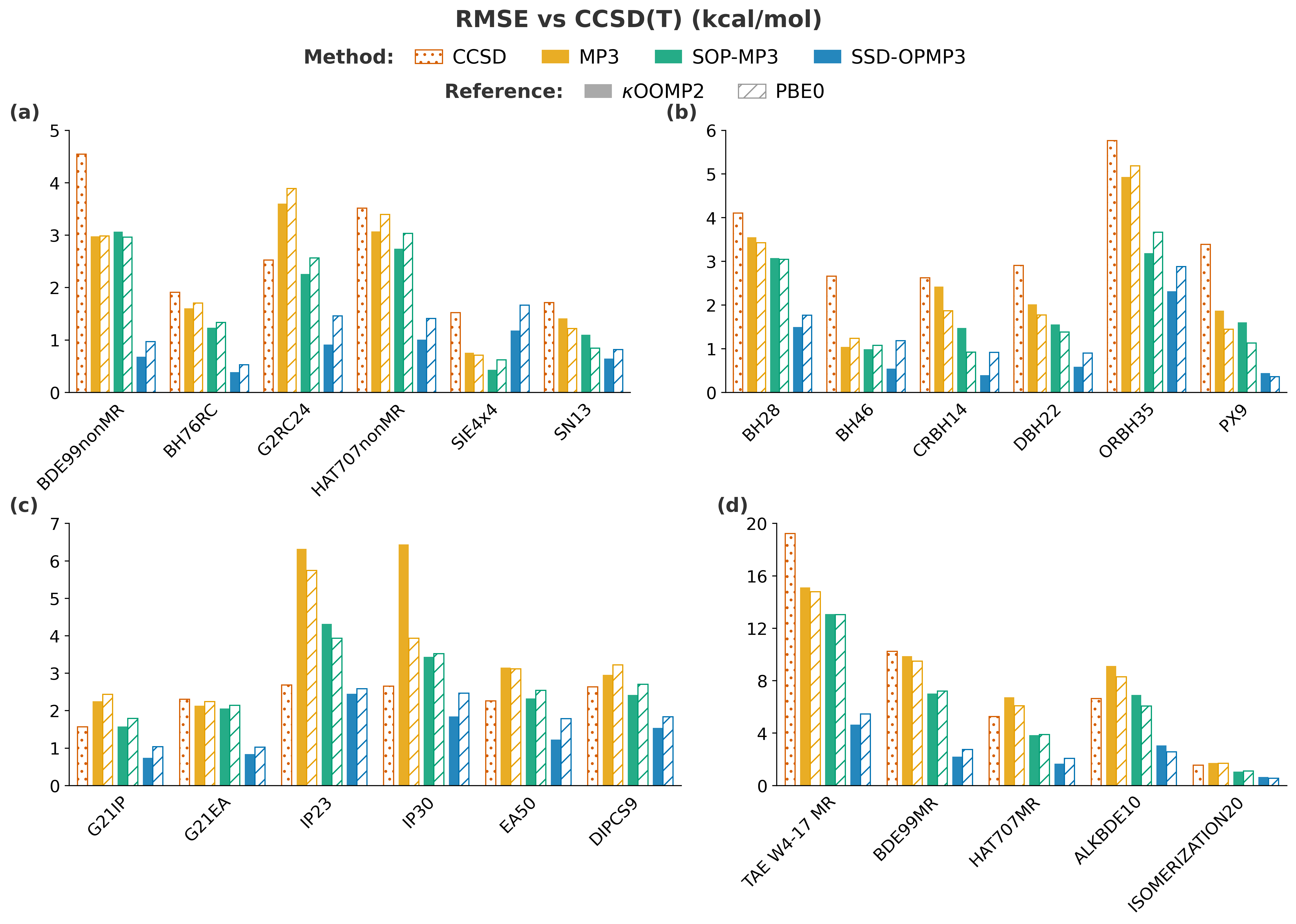}
\caption{Performance of CCSD and the various MP models against CCSD(T) in the same basis set for (a) charge-conserving single-reference thermochemistry, (b) barrier heights, (c) ionization energies and electron affinities, and (d) thermochemistry involving multireference molecules. 
All results are in kcal/mol. Solid bars denote $\kappa$OOMP2 references for the MP methods, while striped bars correspond to PBE0 references (CCSD uses HF orbitals).}
\label{fig:dataset_errors}
\end{figure}

We next consider the performance of SSD-OPMP3 for a wider range of datasets\cite{liang2025gold,goerigk2017look}. Fig. \ref{fig:dataset_errors} (a) shows performance for the thermochemistry of single reference molecules\cite{karton2011w4,zhao2005multi,zhao2005benchmark,curtiss1991gaussian,goerigk2017look}. SSD-OPMP3 with $\kappa$OOMP2 references outperforms CCSD throughout and has an RMSE at or below 1 kcal/mol for all datasets except SIE4$\times$4\cite{goerigk2017look}, although it should be noted that the BDE99nonMR, HAT707nonMR, and SN13 datasets are derived from W4-17 nonMR TAEs. The self-interaction error\cite{perdew1982density,zhang1998challenge,hait2018delocalization} dominated SIE4$\times$4 dataset is a rare case where same-spin damping \textit{hurts} performance, as unscaled SOP-MP3 fares perceptibly better. SSD-OPMP3 with $\kappa$OOMP2 references is nevertheless slightly better than CCSD (RMSE 1.18 kcal/mol vs 1.52 kcal/mol), indicating it to be effective but not perfect. SSD-OPMP3 with PBE0 references fares somewhat worse, but has substantially lower RMSE than CCSD for all datasets save SIE4$\times$4 (RMSE 1.67 kcal/mol). Similar performance is observed for barrier heights\cite{karton2019highly,zhao2005multi,zhao2005benchmark,yu2015reaction,zheng2007representative,chan2018barriometry,karton2012determination}, as depicted in Fig. \ref{fig:dataset_errors} (b). The largest errors arise for the ORBH35 dataset\cite{chan2018barriometry}, which contains reactions involving oxygen and is described by Ref.~\citenum{liang2025gold} to be ``very challenging". SSD-OPMP3 nevertheless significantly improves upon CCSD, traditional MP3, and unscaled SOP-MP3 with both $\kappa$OOMP2 and PBE0 references.  

The performance for molecular ionization energy and electron affinity datasets\cite{curtiss1991gaussian,marie2024reference,ranasinghe2019vertical,ermis2021state} is reported in Fig. \ref{fig:dataset_errors} (c). These prove to be a particularly challenging regime, as traditional MP3 often fares substantially worse than CCSD. SSD-OPMP3 with $\kappa$OOMP2 references nonetheless achieves a lower RMSE than CCSD in all cases, albeit only marginally for the IP23 dataset\cite{marie2024reference} (which contains several multireference species). Same-spin damping uniformly improves performance for both references across all datasets considered. The good performance for these datasets and Pol130 indicates promise towards utilizing Pad{\'e} resummations for studying electronic excited states. 

Fig. \ref{fig:dataset_errors} (d) reports performance for cases with challenging electron correlation\cite{karton2011w4,karton2017w4,yu2015components}, which is made quite evident by the range of the y-axis. SSD-OPMP3 yields excellent performance for the datasets derived from the 17 MR molecules\cite{karton2011w4} in W4-17 (TAE W4-17 MR, BDE99MR, HAT707MR, ISOMERIZATION20), when compared to CCSD, traditional MP3 and SOP-MP3. The $\sim$ 5 kcal/mol RMSE for the TAE 
is nonetheless substantially larger than what was observed for the nonMR subset of W4-17 despite the latter having larger molecules and thus larger TAE. This behavior may be indicative of potential limitations of SSD-OPMP3 for genuinely multireference problems. Finally, ALKBDE10\cite{yu2015components} is an atypical dataset where the PBE0 reference fares slightly better than $\kappa$OOMP2 for SSD-OPMP3, though both references yield much lower errors than other reported methods.

\begin{figure}[htb!]
    \centering
    \includegraphics[width=0.7\linewidth]{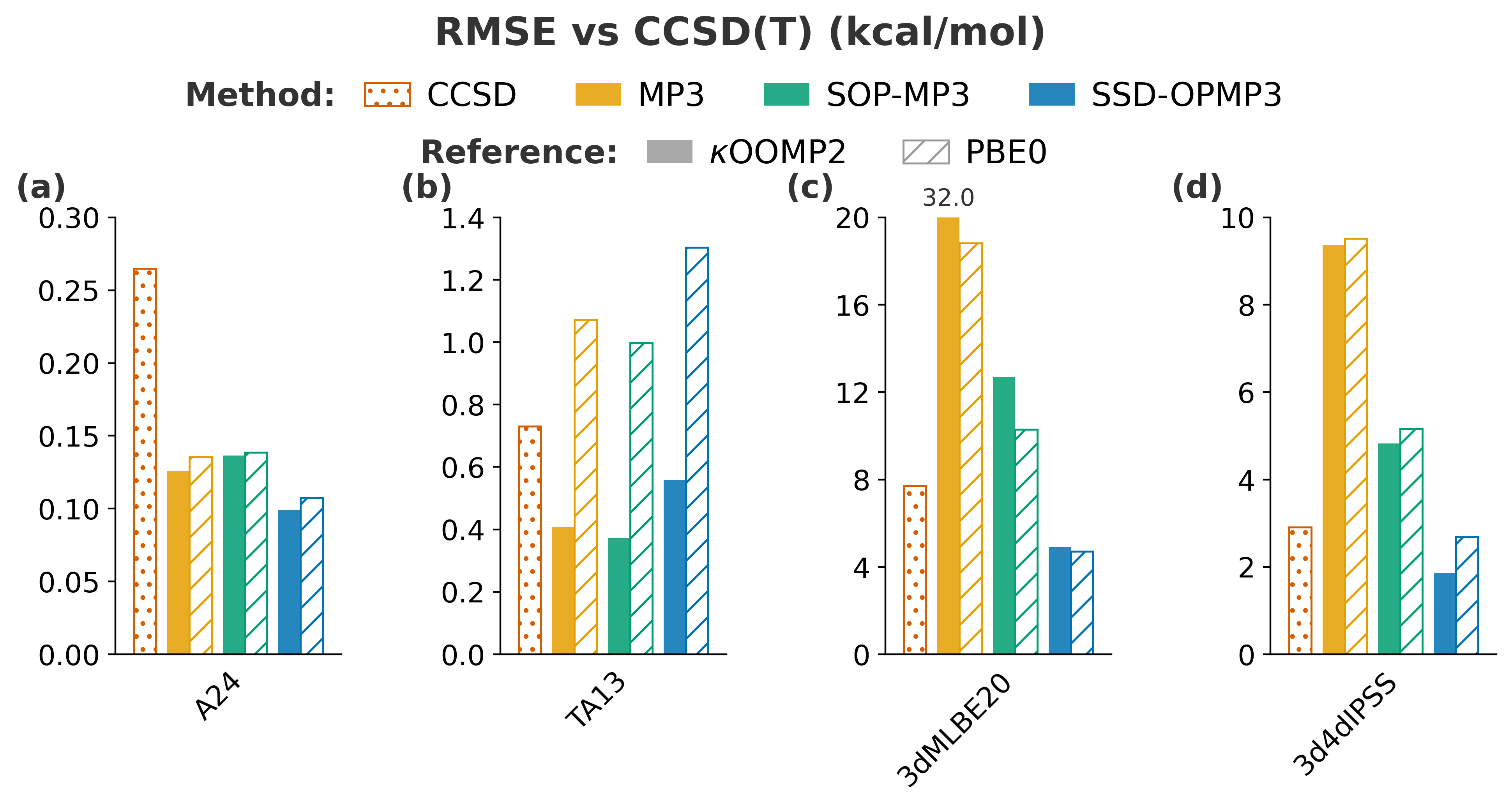}
     \caption{Performance of CCSD and the various MP models against CCSD(T) for (a) A24  and (b) TA13 benchmark sets of noncovalent interactions, as well as the (c) 3dMLBE20 and (d) 3d4dIPSS benchmark sets of  transition metal chemistry. All results are in kcal/mol. Solid bars denote $\kappa$OOMP2 references for the MP methods, while striped bars correspond to PBE0 references (CCSD uses HF orbitals). The large MP3 error with $\kappa$OOMP2 in (c) leads to clipping of axes at 20 kcal/mol, with the true RMSE reported above the bar.}
    \label{fig:ncietTM}
\end{figure}

We finally consider the performance for four additional datasets in Fig. \ref{fig:ncietTM}. A24\cite{rezac2013describing} is composed of noncovalent interactions between small closed-shell molecules where traditional MP3 has half the RMSE of CCSD. SOP-MP3 however has slightly larger errors than MP3, which are reduced by SSD-OPMP3 to make the latter the most accurate protocol, but only marginally so. It is worth comparing this behavior to results for the S22 dataset of larger molecules [Fig. \ref{fig:pair_errors}(d)], where SSD-OPMP3 leads to more significant improvement in performance. TA13\cite{tentscher2013binding} is an extremely challenging dataset of noncovalent interactions between radicals and closed-shell molecules that is particularly sensitive to self-interaction errors. The performance of PBE0 references is quite suboptimal while $\kappa$OOMP2 references lead to better results than CCSD. Interestingly, same-spin damping perceptibly degrades performance compared to SOP-MP3, similar to the SIE4$\times$4 dataset. There is thus some indication that same-spin damping is less effective for datasets with self-interaction error, even for $\kappa$OOMP2 references.

Fig. \ref{fig:ncietTM} also reports performance for the transition metal datasets 3dMLBE20\cite{xu2015practical} (bond dissociation energies of 20 diatomic molecules with one 3d transition metal atom and a ligand) and 3d4dIPSS (ionization energies and same-spin excitation energies of 3d and 4d transition metals). While bond dissociation energies of transition metal diatomics are often extremely challenging for quantum chemical methods\cite{hait2019levels,shee2019achieving}, CCSD(T) has been shown to be effective for this particular subset\cite{fang2017prediction,cheng2017bond} and thus appears to be an acceptable benchmark for assessing the performance of the MP methods studied in this work. We find traditional MP3 to be extremely ineffective, but SOP-MP3 significantly improves upon performance, as shown in Fig. \ref{fig:ncietTM} (c). SSD-OPMP3 improves performance further, with RMSE $\sim$ 5 kcal/mol with both $\kappa$OOMP2 and PBE0 references, surpassing the corresponding value of 7.7 kcal/mol for CCSD. Similar behavior is observed for the ionization energies and same-spin excitation energies of 3d and 4d transition metal atoms, as shown in Fig. \ref{fig:ncietTM} (d). While the SSD-OPMP3 errors overall remain relatively large for the two transition metal datasets, the utility of resummation and same-spin damping is quite apparent and potentially suggests new routes towards simulating transition-metal chemistry.

The general utility of same-spin damping indicates its ability to account for higher-order effects not captured by MP3, the most significant of which ought to be the extent to which $\mathbf{W}_{\sigma\sigma}^{\text{SS}(4)}$ deviates from the ideal geometric series form of $\mathbf{W}_{\sigma\sigma}^{\text{SS}(3)}\left(\mathbf{W}_{\sigma\sigma}^{\text{SS}(2)}\right)^{-1}\mathbf{W}_{\sigma\sigma}^{\text{SS}(3)}$ assumed by SOP-MP3. 
This is similar to the interpretation of SCS-MP2 presented in Ref. \citenum{szabados2006theoretical}, where the scaling parameters $p_{ss}$ and $p_{os}$ are viewed as system-independent approximations to $\left(1-\dfrac{E_c^{\text{SS}(3)}}{E_c^{\text{SS}(2)}}\right)^{-1}$ and $\left(1-\dfrac{E_c^{\text{OS}(3)}}{E_c^{\text{OS}(2)}}\right)^{-1}$ and the correlation energy $p_{ss} E_c^{\text{SS}(2)}+p_{os} E_c^{\text{OS}(2)}$ thus approximately corresponds to a spin-separated scalar Pad{\'e} resummation [cf. Eqn. \eqref{eq:scalar_pade}]. Specifically, if we assume for general chemical systems, it is approximately the case that:
\begin{align}
&\mathbf{W}_{\sigma\sigma}^{\text{SS}(4)}  \approx \left(1-\dfrac{1}{x_{ss}}\right)\mathbf{W}_{\sigma\sigma}^{\text{SS}(3)} +\mathbf{W}_{\sigma\sigma}^{\text{SS}(3)}\left(\mathbf{W}_{\sigma\sigma}^{\text{SS}(2)}\right)^{-1}\mathbf{W}_{\sigma\sigma}^{\text{SS}(3)}
\end{align}
then the fourth-order inclusive, Pad{\'e} resummed, same-spin $\mathbf{W}^\text{SS}$ matrix is:
\begin{align}
\mathbf{W}^{\text{OP-MP4(SS)}}_{\sigma\sigma} &= \mathcal{P}\left(\mathbf{W}_{\sigma\sigma}^{\text{SS}(2)},\mathcal{P}\left(\mathbf{W}_{\sigma\sigma}^{\text{SS}(3)},\mathbf{W}_{\sigma\sigma}^{\text{SS}(4)}-\mathbf{W}_{\sigma\sigma}^{\text{SS}(3)}\left(\mathbf{W}_{\sigma\sigma}^{\text{SS}(2)}\right)^{-1}\mathbf{W}_{\sigma\sigma}^{\text{SS}(3)}\right)\right)\\ 
& \approx \mathcal{P}\left(\mathbf{W}_{\sigma\sigma}^{\text{SS}(2)},x_{ss}\mathbf{W}_{\sigma\sigma}^{\text{SS}(3)}\right) \label{eq:justification}
\end{align}
Of course, a single parameter is a rather inflexible approximation and can be insufficient at times (e.g., the incorrect asymptotic behavior of dispersion in SCS-MP2). Conversely, there is no reason for an empirical $x_{ss}$ not to approximate the effects of $\mathbf{W}_{\sigma\sigma}^{\text{SS}(5)}$ etc., as well. Therefore, while Eqn. \eqref{eq:justification} is a plausible explanation for the success of SSD-OPMP3, we do not attempt to explicitly validate it through the evaluation of $\mathbf{W}^{(4)}$.

To summarize, we have presented an approach for size-consistent, orbital-invariant, Pad{\'e} resummation of the M{\o}ller-Plesset series based on partitioning the correlation energy over occupied orbitals and resumming the resulting self-energy operator. We have further shown that resumming the second- and third-order contributions can lead to near chemical accuracy, as long as a high-quality zeroth-order reference is utilized and the third-order same-spin contribution is damped. This SSD-OPMP3 approach proves to be consistently more accurate than CCSD and MP2.$X$ methods while having a computational cost that is comparable to a single CCSD iteration. SSD-OPMP3 is therefore perhaps the most effective \textit{noniterative} $O(N^6)$ scaling method to date for single-reference systems, approaching the chemical accuracy threshold of 1 kcal/mol even for challenging quantities like total atomization energies. The noniterative nature of the method further allows for highly parallelizable implementations. The SSD-OPMP3 method can already be applied to general chemical problems where greater accuracy than DFT might be desirable, but CCSD(T) is not affordable. Efficient implementations that go beyond our pilot code, including those that leverage local correlation\cite{ma2018explicitly,riplinger2013efficient,nagy2018optimization,ye2024periodic}, tensor hypercontraction\cite{hohenstein2012tensor,parrish2012tensor,hohenstein2012communication,yeh2023low} or stochastic methods\cite{sun2026stochastic,sun2026stochasticb}, would be desirable in this regard.

It is worth recognizing that the choice of high quality reference orbitals ($\kappa$OOMP2 or DFT) contributes substantially to accuracy, as previously shown by studies focusing on traditional MP2.$X$ methods\cite{bertels2019third,rettig2020third} and the results presented in Figs.~\ref{fig:w417errors}--\ref{fig:ncietTM} of this work. Resummation and same-spin damping are, however, critical for substantial further improvement in accuracy relative to traditional MP theory for most chemical problems. The sensitivity of the results (and the same-spin damping parameter $x_{ss}$) to the choice of reference method further suggests that self-consistent orbital-optimization could lead to superior results and can potentially be accessed through generalization of orbital-optimized MP3\cite{bozkaya2011orbital,soydas2013assessment}. The resulting approach will have an iterative $O(N^6)$ scaling, and comparisons to similarly scaling methods of greater accuracy than CCSD would be worthwhile \cite{kats2013communication,behnle2021oo,behnle2022uremp,takatani2008improvement,small2020remarkable}. 

Further gains in accuracy may result from the explicit use of $\mathbf{W}^{(4)}$ in the resummation, at $O(N^7)$ cost, or through hybridization with Kohn-Sham DFT. Careful validation against accurate post-CCSD(T) methods like CCSDT(Q)\cite{bomble2005coupled}/CCSDTQ\cite{kucharski1992coupled} would be essential in this regime, in light of the current near-chemical accuracy of SSD-OPMP3. Improved performance may also be realizable through the use of regularized perturbation theory\cite{carter2023repartitioned,wang2026thirdsc} or adiabatic connection\cite{bystrom2026size} in the resummation. These approaches are likely to be necessary for chemical problems with stronger levels of electron correlation than those considered in this work, such as bond dissociation. Work along these directions is currently in progress.

\section*{Computational methods}
All calculations were performed with the Q-Chem\cite{epifanovsky2021software} and PySCF\cite{sun2020recent} packages, with PySCF using libxc~\cite{lehtola2018recent} to evaluate density functionals. Q-Chem was used to generate most of the reference Slater determinants, while the ADC(3) module\cite{banerjee2023algebraic} of PySCF was leveraged for MP calculations leading to $\mathbf{W}$ matrix construction. The codebase for the generation and resummation of $\mathbf{W}$ is shared via a Zenodo repository\cite{bystrom_2026_23482348}. 

The HF/DFT orbitals were optimized under restricted open-shell conditions to prevent spin contamination, while $\kappa$OOMP2 orbitals were optimized with spin-unrestricted orbitals ($\kappa=1.45$). The resulting Slater determinants were processed in the manner described in Ref \citenum{rettig2020third}. In brief, a spin-unrestricted HF Fock matrix is built from the density matrix of the reference determinant. The occupied and virtual blocks of this Fock matrix are independently diagonalized (`semicanonicalized') to obtain orbital energies without mixing the occupied and virtual subspaces. The $\mathbf{W}^{(n)}$ matrices were subsequently generated following Eqns \eqref{eq:wss_def} and \eqref{eq:wos_def}. The DFT orbital energies thus do not explicitly enter the calculations for $\mathbf{W}^{(n)}$ matrices or the total energy. However, the variation in $x_{ss}$ over the tested references indicates that the extent to which electrons of the same-spin interact with each other (via the exchange-correlation term) has a significant impact on MP SS correlation. The non-Brillouin singles energy at the MP2 level is computed from the semicanonicalized Fock matrix and is directly added to the energy, without any resummation or scaling. No other contributions from the non-Brillouin singles (e.g. third-order terms) were considered.  

Most results in this work use the aug-cc-pCVTZ basis without any frozen-core, or density-fitting. If unavailable (such as for K, Ca, and 3d/4d transition metals), aug-cc-pwCVTZ\cite{balabanov2005a,peterson2007energy} was used instead (with the standard PP pseudopotentials for the 4d elements). The bases unavailable by default in Q-Chem were obtained from the Basis Set Exchange\cite{pritchard2019a} or cc-Repo\cite{hill2022ccrepo}. The major exception to this general principle is the S22 dataset, for which we use counterpoise corrected aug-cc-pVTZ  with frozen-core [for which Ref \citenum{burns2014appointing} provides CCSD(T) results]. Density-fitting with the aug-cc-pVQZ-RIFIT auxiliary basis\cite{weigend2002a} was also carried out for the S22 MP calculations. The composition of all datasets utilized in this work (other than 3dMLBE20\cite{xu2015practical}) corresponds to Ref \citenum{liang2025gold}, as do the geometries and spin states. The excitation energy of neutral atomic Cu in 3d4dIPSS was not considered, as the two primary sources referenced by Ref. \citenum{liang2025gold} disagree about the nature of the excitation ($3d\to4s$ in Ref. \citenum{balabanov2005a} and $3d\to4p$ in Ref. \citenum{luo2014density}).

Dipole moments and static polarizabilities were found using finite difference with applied external electric fields, as in Ref. \citenum{liang2025gold}. We also note that while the Pol130 dataset contains 130 molecules, we only utilize 370 diagonal components of the static polarizability tensor. The 20 missing components correspond to the elements perpendicular to the bond axis for nine linear radicals (OH, OF, OCl, SH, SF, SCl, PS, NO and NCO) and the components perpendicular to the C-O bond axis for the related \ce{CH3O.} radical. This choice stemmed from the inability of real orbitals to describe the cylindrical symmetry of the electron density for $^2\Pi$ systems, which affects polarizability calculations with approximate electronic structure\cite{brakestad2020static}. 

We explored a variety of resummation schemes aside from Pad{\'e}, including Pad{\'e}-Borel\cite{zhao2024meijer} and Landau-Zener\cite{sparpaglione1988dielectric}. The alternative approaches were found to yield inferior results and are thus not discussed in detail. 

\section*{Data Availability Statement}
The data underlying this study are openly available in Zenodo at \url{https://doi.org/10.5281/zenodo.23482348}. The repository\cite{bystrom_2026_23482348} contains PySCF code for $\mathbf{W}$ generation, code for performing resummation, computed reaction energies, error statistics, interactive HTMLs for data visualization, and raw data ($\mathbf{W}$ matrices, associated molden files, etc.).

\section*{Acknowledgments}
The Flatiron Institute is a division of the Simons Foundation. 
\bibliography{references}
\end{document}